# Niobium doping induced mirror twin boundaries in MBE grown WSe$_2$ monolayers

Bo Wang[1,2], Yipu Xia[3], Junqiu Zhang[3], Hannu-Pekka Komsa[4,*], Maohai Xie[3], Yong Peng[1], and Chuanhong Jin[2,1,5, *]

[1] *Key Laboratory for Magnetism and Magnetic Materials of Ministry of Education, School of Physical Science and Technology, Lanzhou University, Lanzhou 730000, China*
[2] *State Key Laboratory of Silicon Materials, School of Materials Science and Engineering, Zhejiang University, Hangzhou 310024, China*
[3] *Physics Department, The University of Hong Kong, Pokfulam Road, Hong Kong 999077 , China*
[4] *Department of Applied Physics, Aalto University, 00076 Aalto, Finland*
[5] *Hunan Institute of Advanced Sensing and Information Technology, Xiangtan University, Xiangtan 411201, China*
*Email: Hannu-Pekka Komsa, hannu-pekka.komsa@aalto.fi; Chuanhong Jin, chhjin@zju.edu.cn

## ABSTRACT

Mirror twin boundary (MTB) brings unique 1D physics and properties into two-dimensional transition metal dichalcogenides (TMDCs), but they were rarely observed in non-Mo-based TMDCs. Herein, by post-growth Nb doping, high density 4|4E-W and 4|4P-Se MTBs were introduced into molecular beam epitaxy (MBE) grown WSe$_2$ monolayers. Of them, 4|4E-W MTB with a novel structure was discovered experimentally for the first time, while 4|4P-Se MTBs present a random permutations of W and Nb, forming a 1D alloy system. Comparison between the doped and non-doped WSe$_2$ confirmed that Nb dopants are essential for MTB formation. Furthermore, quantitative statistics reveal the areal density of MTBs is directly proportional to the concentration of Nb dopants. To unravel the injection pathway of Nb dopants, first-principles calculations about a set of formation energies for excess Nb atoms with different configurations were conducted, based on which a model explaining the origin of MTBs introduced by excess metal was built. We conclude that the formation of MTBs is mainly driven by the collective evolution of excess Nb atoms introduced into the lattice of host WSe$_2$ crystal and subsequent displacement of metal atoms (W or Nb). This study provides a novel way to tailor the MTBs in 2D TMDC materials via proper metal doping and presents a new opportunities for exploring the intriguing properties.

## 1 Introduction

Two-dimensional (2D) materials, such as graphene, hexagonal boron nitride, and transition-metal dichalcogenides (TMDCs) hold great promise for applications in nanoelectronics[1], optoelectronics[2], valleytronics[3], and catalysis[4] owing to their reduced dimensionality and unique structures and properties. It is well known in semiconductor industry that doping and defect engineering are among the two most reliable routes for tuning the structural and electronic properties of materials and even adding new functionalities to them. These approaches should also work for 2D semiconductors such as TMDCs. As demonstrated previously, introduction of suitable dopants and/or defects into 2D TMDCs induced tunable magnetism[5,6], controllable charge density wave[7], engineerable bandgap and carrier mobility[8,9], towards the device application[10,11].

Mirror twin boundaries (MTBs), also called inversion domain boundaries (IDBs), are a kind of one-dimensional (1D) defects between two grains rotated by 60º (thus mirrored with respect to each other). Such 1D MTBs in 2D TMDCs can host a variety of novel 1D physics that include Tomonaga-Luttinger liquid behavior[12-14], Peierls instability[15], and quantum confinement effect[13]. This has attracted particular attention to realize controllable engineering of MTBs, especially in molybdenum dichalcogenides. For example, Coelho et al[16] reported that MTBs in MBE grown MoSe$_2$ and MoTe$_2$ monolayers could be controlled via excess molybdenum doping. Jiao et al[17] found that the MTBs in MoSe$_2$ could be eliminated by vacuum annealing. So far, the MTB networks with a high areal density were only observed in MBE grown Mo-based TMDCs, such as MoSe$_2$[18] and MoTe$_2$[19], but rarely in other TMDCs materials [20]. Generally, the route to introducing the 1D MTBs into non-Mo-based TMDCs such as WSe$_2$ remains unknown.

As a highly promising material for valleytronics,[21-23] single layer WSe$_2$ and other W based dichalcogenides have not yet been reported to contain intrinsic MTBs due to the higher formation energies of MTBs as compared to MoTe$_2$ and MoSe$_2$.[24] At present, the only operational approach was utilizing high energy electron beam irradiations as the stimulus to enable the formation of 55|8 type MTB in WSe$_2$,[25] although this method clearly cannot be scaled up for large-area fabrication. Herein, we report a novel approach to prepare high density MTB networks in MBE grown WSe$_2$ monolayers that relies on excess niobium (Nb) metal doping into the host 2D materials. As characterized by atomic resolution annular-dark-field scanning transmission electron microscopy (ADF-STEM), two MTB configurations were found in Nb-doped WSe$_2$ monolayer: 4|4E-W, a new type of MTB, and 4|4P-Se MTBs with Nb dopants preferably concentrating on them. Nb dopants were confirmed as a necessity for MTB formation via careful comparison of the structures of doped and non-doped sample regions, and the areal density of MTBs was found to be proportional to the Nb concentration. Quantitative relationship between Nb dopants and MTBs was built based on the statistical data by systemically analyzing several regions with different Nb concentrations. To unravel the injection pathway of Nb dopants, the first-principles calculations were further carried out, based on which an atomic-level model is built to explain the microscopic mechanisms for formation of the MTB loops.

## 2 Results

**2.1 Overview of the Nb-doped and undoped WSe$_2$ flakes**

Nb-doped WSe$_2$ samples were characterized by atomic resolution ADF-STEM, a Z-contrast imaging technique[26] that gives I~Z$^{1.6\sim2.0}$, where I and Z are the intensity and the effective atomic number/thickness, respectively. Considering the constituent lattice elements in our samples, i.e., $Z_{Nb}$=41, $Z_W$=74, and $Z_{Se}$=34, those lattice atoms can thus be unambiguously distinguished in single layer Nb-doped WSe$_2$ via ADF-STEM. Hence, the dimmer contrast atoms extensively distributed on a large-scale sample region shown in Fig. 1 (a) are assigned to Nb dopants, which was further supported by chemical analysis via EDS (Fig. S1). As seen, majority of the host WSe$_2$ sample is doped with Nb along with small undoped region, indicating an uneven Nb distribution across the sample. Fast Fourier transform spectrum (Fig. 1(c)) of Fig. 1 (a) shows that the diffraction patterns of WSe$_2$ are inter-linked to form a so-called David star, a characteristic of the MTB networks.[10] To map out the spatial distribution of MTB network, Fig. 1 (a) is then low-pass filtered, after which a network of 1D dimmer lines clearly appear as shown in Fig. 1 (b), indicating a high density MTB network. Those MTBs are all oriented along the [100], [010], and [110] directions, i. e., the zigzag directions, and they further form triangular loops with noticeable inward and outward kinks in their edges, as depicted in Fig.1 (b). Careful comparison between Figs. 1 (b) and (a) reveals that the MTB network formed only in the Nb-doped region, but not in pure WSe$_2$ or NbSe$_2$ areas (refer to Fig. 1 (d) and Fig. S2 for the atomically resolved lattices of WSe$_2$ and NbSe$_2$, respectively). This suggests that the induced MTBs in WSe$_2$ must be correlated with the Nb doping.

To further study the influence of Nb dopants on the host lattice, selected area electron diffraction (SAED) was conducted as shown in Fig. 1 (e). The lattice constant of Nb-doped WSe$_2$ was measured as 3.29(4) Å, a value almost identical to that of WSe$_2$ (3.29Å). In general, lattice constant of TMDC alloys follow fairly closely the Vegard's law [27, 28] which can be written as

$$a_{Nb_xW_{(1-x)}Se_2} = xa_{NbSe_2} + (1-x)a_{WSe_2} \qquad (1)$$

where $a_{Nb_xW_{(1-x)}Se_2}$, $a_{NbSe_2}$, and $a_{WSe_2}$ are the lattice constant of alloy Nb$_x$W$_{(1-x)}$Se$_2$, NbSe$_2$ (3.45 Å), and WSe$_2$ (3.29 Å), respectively; and *x* is the doping concentration of Nb. Eq. (1) shows that lattice constant of Nb-doped WSe$_2$ should increase monotonically with doping concentration, and for a typical concentration of *x*=10 % in our samples, Eq. (1) gives $a_{Nb_{0.1}W_{0.9}Se_2}$=3.306 Å, which is clearly different from the experimental

value of 3.29(4) Å. This comparison suggest that lattice distortion (and thus strain) caused by Nb doping is

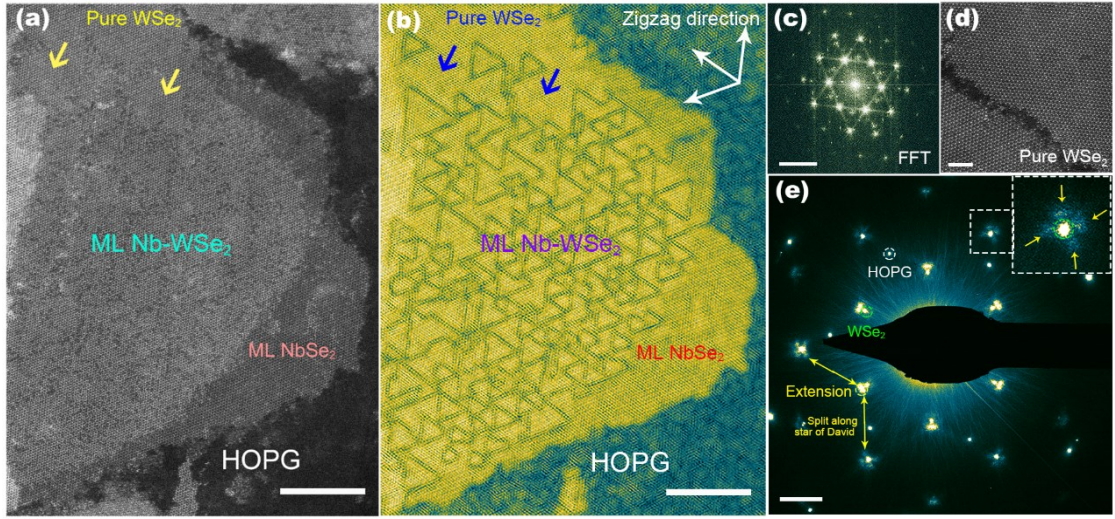

**Figure 1** Overview of Nb doped WSe$_2$ monolayers. (a) An atomic-scale ADF-STEM image of Nb-doped WSe$_2$ sample, where dimmer contrast randomly distributed in WSe$_2$ layer corresponds to Nb dopants. Scale bar, 10 nm. (b) Low pass filtered ADF-STEM image showing a network of dimmer lines, i.e., MTBs in Nb-doped sample area. Scale bar, 10 nm (c) A FFT spectrum of (a) showing the existence of a Star of David, a strong signal of 1D system. Scale bar, 5 nm$^{-1}$. (d) A zoom-in ADF-STEM image of undoped WSe$_2$. Scale bar, 2 nm. (e) A SAED pattern collected from the Nb doping area, in which extension (pointed by yellow arrows) appear around WSe$_2$ pattern, which is more clearly shown in inset. Scale bar, 2 nm$^{-1}$.

mainly localized around the MTBs. Notably, a new symmetry was also introduced as evidenced by the unusual extensions appearing around WSe$_2$ pattern and split along the Star of David (yellow arrows in Fig. 1 (e) and more clearly in Fig. S2), indicating that high density MTB network in WSe$_2$ possess a quasi-periodic order, similar to the MoSe$_2$ samples reported previously.[19, 29]

**2.2 Atomic structure of mirror twin boundaries in Nb-doped WSe$_2$**

In order to clarify the exact relationship between Nb dopants and MTBs, atomic resolution ADF-STEM images were analyzed. The ADF-STEM image in Fig. 2 (a) indicates that there exist two types of MTBs in the Nb-doped WSe$_2$ sample. The dominant MTB is 4|4P-Se (Fig. 2 (b) right panel), which has the same structure as those MTBs reported for MoSe$_2$ and MoTe$_2$[24], in which tetragons (4-member rings, the basic units marked by gray tetragons) are lined up sharing a point at Se$_2$ sites and with a Se-termination of WSe$_2$ half-lattices (along the Se-zigzag (Se-ZZ) direction). Surprisingly, we also found a new MTB, here denoted as 4|4E-W (Fig. 2 (b) left panel), which has not been reported before. Inside the 4|4E-W MTB, tetragons share the edge at W-Se bonds and with a W-termination of WSe$_2$ half-lattices (along W-zigzag (W-ZZ) direction), as confirmed by intensity analysis (see pink profiles in Fig. 2 (b)). Compared to 4|4P-Se, 4|4E-W MTBs are rarely found and shorter in length in the Nb doped WSe$_2$ sample as seen in Fig. 2 (a).

Even though 4|4E-W MTBs are located in Nb-doped region, no Nb dopants were observed at these MTBs, as shown in Fig. 2 (d). Hence, from here on, we mainly focus on the Nb dopants in 4|4P-Se MTBs. Fig. 2 (c) presents a close-up view of a single 4|4P-Se MTB loop (Nb dopants were marked by green circles) whose three edges (corresponding to three MTBs) are numbered as I, II, and III, respectively. Their atomic structures and Nb concentrations (I: 21.4%, II: 35.7%, III: 32.1%) are also shown in Fig. 2 (d). As seen, Nb dopants substitute W atoms randomly at 4|4P-Se MTBs with no obvious clustering. Note that symmetry breaking due to the random permutations of W and Nb along the MTB leads to a 1D random alloy system.

Two configuration of isolated Nb dopants in WSe$_2$ crystal were observed, namely Nb$_W$ (Nb occupying a W site, marked by green circle in Fig. 2 (e)) and h$_{Nb}$ (Nb occupying a hollow site, marked by blue circle in

Fig.2 (e)). Notably, the relative proportion of $Nb_w$ and $h_{Nb}$ defects is 99.7 % and 0.3 %, respectively.

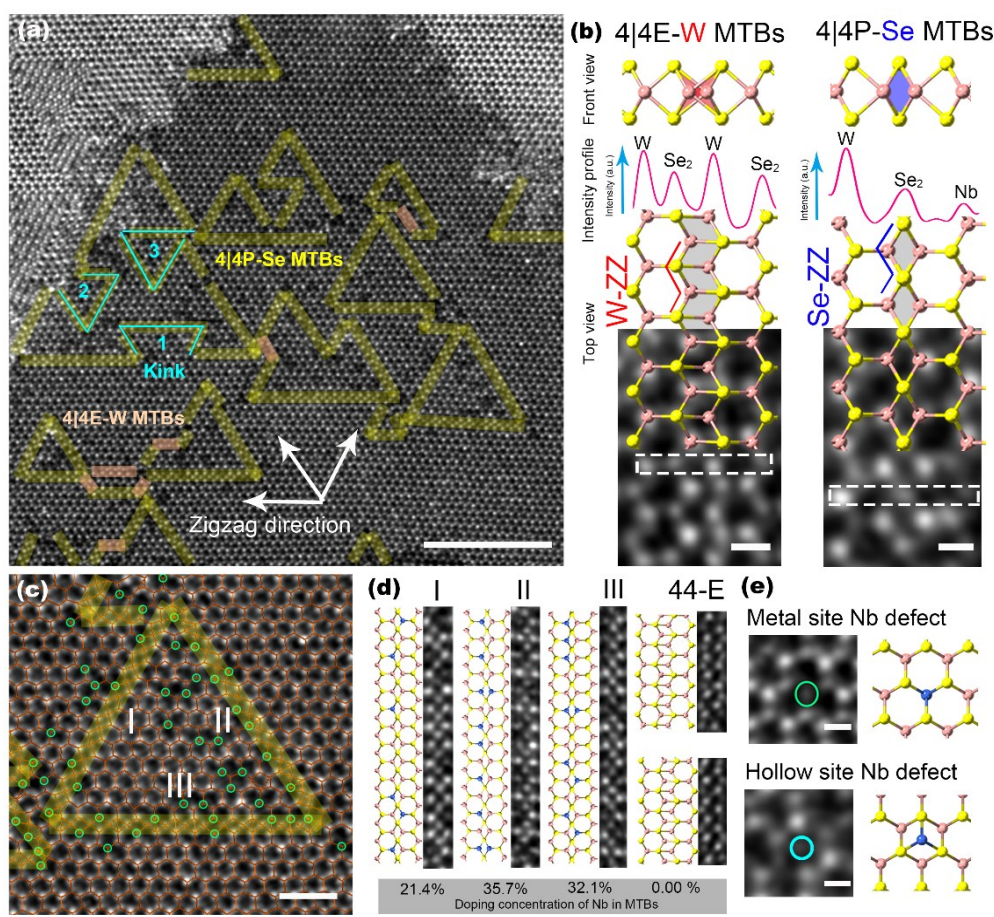

**Figure 2** MTB network. (a) An atomic-scale ADF-STEM image of Nb doping area, where 4|4P-Se and 4|4E-W MTBs are marked by yellow and orange bands, respectively. Three kinks in MTB loops were marked by blue line and namely 1, 2, 3, respectively. Scale bar, 5 nm. (b) Atomic structures and ADF-STEM images of 4|4P-Se MTBs and 4|4E-W MTBs. Intensity profiles shown at the top are collected along the white dotted lines. Scale bar, 0.2 nm. (c) A Zoom-in image of MTB loop, where green circles indicate Nb dopants. Scale bar, 1 nm. (d) Atomic structure of 4|4-Se MTBs marked in (c) and two 4|4E-W MTBs collected in figure S3. (e) Metal site Nb defect and Hollow site Nb dopant (left) and their atomic structures (right). Scale bar, 0.2 nm.

Analysis of the spatial distribution of Nb dopants in Fig. 2 (a) reveals that around 67.9% of Nb dopants are preferably distributed inside the $WSe_2$ sample regions enclosed by the MTB loop, leading to a nonuniform distribution of Nb dopants, which may stem from the growth history of MTBs as will be discussed later on.

**2.3 Relationship between MTBs and Nb dopants**

Statistical analysis was further conducted to unravel the relationship between MTBs and Nb dopants and clarify the origin of these MTBs. The results are summarized in Tab. 1, for which data from three sample areas (Fig. S4) with different Nb concentration i.e., 6.7%, 8.9%, and 12.8 %, respectively, were analyzed. For brevity, the number of tetragons (as mentioned above) was counted to represent the total length of MTBs. Our analysis shows that the ratios of the number of tetragons to that of Nb dopants are around 1.63:1 in all three cases, irrespective of the areal Nb concentrations, or in other words, the density of MTB network is proportional to the areal Nb concentration. In addition, the areal Nb concentration is lower than the local Nb concentration in MTBs as summarized in Tab. 1, which further confirms that Nb dopants are concentrated on the MTBs. For instance, the concentration of Nb in MTBs (20.1%) is 3 times larger than in whole area (6.7%). In all three cases, concentrations of Nb in MTBs are stable at around 20%~24% and slowly increase with the areal Nb concentration. Thus, under a relatively low Nb dopant concentration, most of the Nb dopants are confined to 1D MTBs.

Table 1 Statistical analyses from different niobium concentration areas

| Areal Nb concentration | 6.7% | 8.9% | 12.8% |
|---|---|---|---|
| Local Nb concentration in MTB | 20.1% | 20.4% | 24.1% |
| Tetragon: Nb dopant | 1.647 : 1 | 1.624:1 | 1.637:1 |

**2.4 Formation of MTBs in Nb-doped WSe$_2$**

To understand the roles played by Nb atoms in the formation of MTBs in WSe$_2$ monolayer, we need to clarify the interaction between the excess Nb and host WSe$_2$. Thus, we calculated the formation energies of additional Nb atoms in different configurations to explain how Nb enters WSe$_2$ lattice during the MBE process. Formation energies for all configurations with additional Nb atoms are defined as

$$E_f = E(defect) - [E(pristine) + n_{Nb}\mu_{Nb}] \qquad (2)$$

where *E(defect)* and *E(pristine)* are the total energies of the supercell with the Nb dopants and pristine WSe$_2$ layer, respectively, $n_{Nb}$ is the number of Nb dopants and $\mu_{Nb}$ is the chemical potential of the Nb atom in the isolated Nb$_2$ dimer.

The following defect configurations were considered: Nb adatom on top of W (adatom), Nb at interstitial site (interstitial), Nb substituting for Se with Se pushed to adatom site (Se-sub), Nb substituting W with W pushed to interstitial site (Nb$_w$+W(interstitial)), and Nb substituting W with W pushed to Se site and Se to adatom site(Nb$_w$+W(Se-sub)). The final optimized geometries are displayed in Fig. 3 (a-f) and the corresponding formation energies are also shown. The Se-sub configuration has the lowest energy, exactly as found in the case of MoSe$_2$.[30] Pure interstitial site has the highest energy, even higher than adatom. Instead of going to interstitial site, Nb would rather enter W site and displace W to interstitial site. The preference for Nb to substitute W is probably related to stability of NbSe$_2$ over WSe$_2$ and even MoSe$_2$. The calculated heats of formation are -1.66 eV, -2.01 eV, and -2.50 eV for WSe$_2$, MoSe$_2$ and NbSe$_2$, respectively. Comparing the energy of adatom (0.53 eV) with other configurations (<0 eV, except interstitial 0.9eV), WSe$_2$ monolayer would catch those excess Nb atoms and keep them inside the lattice rather than let the adatoms cluster on the lattice surface. This tendency will create local metal-rich conditions inside the host WSe$_2$ lattice.

We also consider MTB loop structures of different sizes. Their formation requires introduction of additional W or Nb atoms. Here, we treat all additional metal atoms as Nb for simplicity, and thus we can still use Eq. (2). The Nb atoms are equidistantly and randomly placed at the in- or outside of the MTB. One example is shown in Fig. 3 (f). Furthermore, we compare these to formation energies of MTB loops where the number of Nb atoms is different from the size of loop, which requires choosing chemical potential for W:

$$E_f = E(defect) - [E(pristine) + n_{Nb}\mu_{Nb} + n_W\mu_W] \qquad (3)$$

Consequently the results depend on the adopted choice of $\mu_{Nb}$ and $\mu_W$, and thus care is needed when comparing the numbers. $\mu_W$ is from W$_2$ dimer, which is consistent with the choice for Nb. We note that fairly similar results are obtained if we choose isolated Nb and W atoms or substitutional Nb$_W$ defect. The formation energies as a function of the number of added (Nb) atoms are shown in Fig. 3 (g). Comparison of the isolated defect energies and "N Nb MTB" shows that for N > 2, Nb-stabilized MTBs possesses clearly the lowest formation energy, indicating that the MTB is the energy-preferable structure under the metal-rich condition. Furthermore, the Nb-stabilized MTBs are significantly lower in energy than the MTBs containing only W (with or without addition of Nb$_W$ defects far from it).

Based on the calculated and experimental results, we can now propose a qualitative atomic-scale pathway for the formation of MTBs as triggered by excess Nb doping, summarized in Fig. 4, where four different pathways (marked by purple) for Nb entering WSe$_2$ lattice are considered, that is Se-sub, interstitial,

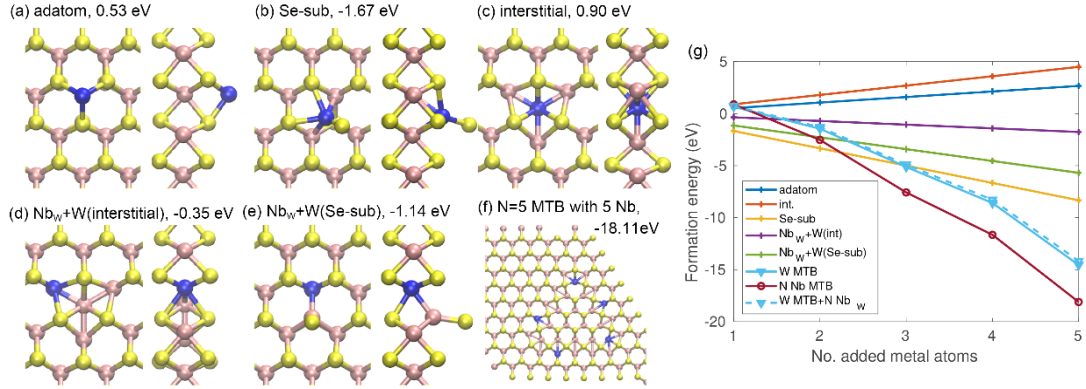

**Figure 3** (a-f) Optimized defect geometries. (g) Energies as a function of the number of added Nb/W metal atoms $N$ for $N$ isolated Nb-defects compared to size-$N$ MTBs with 0 or $N$ Nb atoms placed at MTBs.

Nb$_W$+W(interstitial) and Nb$_W$+W(Se-Sub). The number of Nb dopant is set as 3 for all pathway due to the energy calculation in Fig. 3 (g).

For Nb doping at interstitial site shown in Fig. 4 (a) left panel, the neighboring Nb/W atoms can be pushed to the centers of hexagons (black arrows) and form an N=3 4|4P-Se MTBs with 3Nb (Fig. 4 (a) right panel). Although the pathway in Fig. 4 (a) has been used to explain the formation of MTBs in MoSe$_2$ and MoTe$_2$, interstitial site dopants have the highest formation energy as mentioned before (Fig. 3 (c)), which means this pathway is not energy-preferable in Nb-doped WSe$_2$. For Nb doping at other sites with lower formation energies, the corresponding MTB transform mechanism is shown in Fig. 4(b). Se-Sub and Nb$_W$+W(Se-Sub) dopants will firstly transform to Nb$_W$+W (interstitial) which has a similar structure to that of interstitial site configuration but a lower energy; Nb$_W$+W (interstitial) could also form MTB loop from the transform 4|4E-W MTBs (black arrows in Fig. 4 (b) middle). After the formation of MTB loop, Nb dopants will stay at W site when Nb dopants enter into WSe$_2$ lattice along four pathway mentioned previously, agree well with the statistical results that 99.7% Nb dopants are Nb$_W$ defects in experiment. We note that 4|4E-W MTB (marked by red dot rectangle, Fig. 4 (a) middle panel) can be found as an intermediate product during the formation of 4|4P-Se MTB loops and the transformation can occur without adding or losing any excess atoms, as shown in Fig. 4 (a) from middle panel to right panel.

Following the formation of N=3 4|4P-Se MTB loop, the further excess metal atoms arriving on the surface can help the subsequent growth of the 4|4P-Se MTB loop as sketched in Fig. 4(c). Thus we denote the N=3 MTB loops in Fig. 4 (a) and (b) as MTBs seeds. Due to the unknown doping site of Nb in MBE process, to simplify the growth model, excess Nb dopants are considered to be of Nb$_W$+W(interstitial) type and the transformation processes via Se-Sub/Nb$_W$+W(Se-Sub) to Nb$_W$+W(interstitial) are ignored. In addition, the type of spatial doping sites were classified as outside doping and interior doping. When an Nb atom is added at the outside of MTB loop shown as 3Nb 3MTB+1Nb in Fig. 4 (c) left panel, MTB loop will grow larger and the Nb dopants that were previously at the 4|4P-Se MTB may be locked in the interior of the enlarged MTB loop. This process could explain the nonuniform distribution of Nb dopants shown previously. In the situation where Nb atoms are added at the inside of MTB loop shown as 4Nb 4MTB+1Nb in Fig. 4 (c), an inward kink of MTB loop would form as marked by blue line.

Depending on the position of the Nb atoms inside the MTB loop, the kinks can further increase in size, as also observed in our experiment as marked by blue lines in Fig. 2 (a). From 1 to 3, those kinks may show three different instances of new MTB loop formation processes. Generally, the evolution of the MTB network should aim to minimize the number of corners, due to the associated energy penalty, but also to maximize the number of Nb atoms at the 4|4P-Se MTB, due to the associated energy gain. Competition between these two aspects can then lead to the complex networks seen in our experiments.

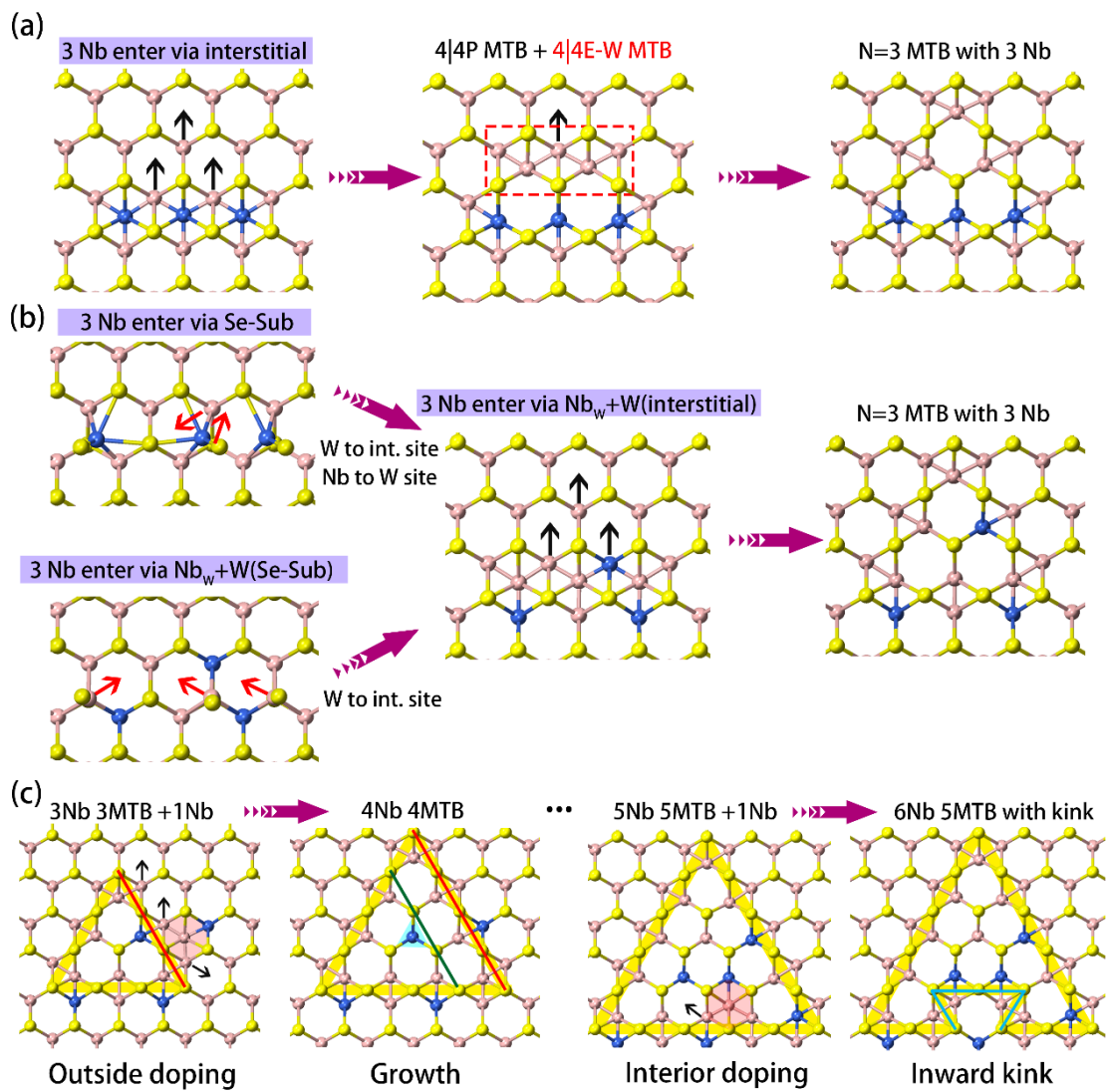

**Figure 4** Formation model of MTB loops. (a) 3 Nb dopants enter into WSe$_2$ lattice along interstitial pathway and transform to N=3 MTB loop, 4|4E-W MTB has been marked by red dotted rectangle, black arrows shows the transformation process of 4|4E-W MTBs. (b) 3 Nb dopants enter into WSe$_2$ lattice along Se-sub, Nb$_W$+W(Se-Sub), Nb$_W$+W(interstitial) pathway and transform to N=3 MTB loop. Red arrows shows how Se-sub, Nb$_W$+W(Se-Sub) transfom to Nb$_W$+W(interstitial)) structure. (c) Two examples MTB growth: one is doping at outside of MTB loop and the other is doping at interior of MTB loop.

During the formation process of MTB loops (Fig. 4 (b) and (c)), the number of tetragons should be directly related to the number of Nb dopants in a 3:1 ratio in all cases, suggesting that the total length of MTB should increase with the number of Nb dopants. However, our statistical analysis yielded the ratio of the number of tetragons to Nb dopants at around 1.63:1, rather than 3:1 in our model, indicating that not all of the Nb dopants contribute to MTB growth. Except the configurations in Fig. 3 (a-f), Nb atoms could also directly replace the W atoms to enter into lattice, which is supported by calculation[31]. Furthermore, calculated results (Fig. S6) prove that Nb atoms prefer to replace the W atoms at 4|4P-Se MTBs or its vicinity. This can lead to two consequences. First, the local metal concentration would not be increased, meaning that this kind of dopants would not contribute to MTBs growth. Second, Nb atoms would be concentrated on 4|4P-Se MTBs, which was observed in experiment and presented in table 1.

A full picture on how the Nb atoms induce the formation of a high density MTB network can finally be drawn. First, WSe$_2$ crystal absorbs the injected Nb atoms into the lattice and creates a local metal rich conditions inside the lattice. Then, the close-by Nb dopants migrate along the pathway shown in Fig. 4 (b) to form the seeds for MTB loops (N=2 or N=3 MTB loops). As the number of Nb dopants further increases, the seeds will grow up in size following the pathway illustrated in Fig. 4 (c). Kinks appearing on edges can also serve as new seeds. During the growth process, most of 4|4E-W MTBs transform into 4|4P-Se MTBs, but a small part of 4|4E-W MTBs could survive since their transformation process will be inhibited by the surroundings, they arise indirectly during the evolution of the MTB network. Meanwhile, some Nb dopants directly substitute W atoms in MTBs and its vicinity, and do not contribute to the MTB growth. Eventually, a high density network of MTBs is formed by excess Nb atoms.

## 3 Discussion

Thus far, the main approach to introduce MTBs is to create a metal-rich conditions, e.g., by using high energy electron beam irradiation[25, 32], changing the MBE conditions[17], and post-doping with excess metal atoms[16]. For the post-doping approach used in this study, an important factor is that excess Nb atoms prefer entering into WSe$_2$ lattice rather than clustering on surface. Thus, to introduce MTB in other TMDC materials, dopants need to enter into lattices easily and form a local metal-rich conditions inside lattice. As systematically studied previously[30], a larger lattices constant *a* of the host TMDC material could reduce the energy for dopants to enter into lattices, and a smaller atomic radius of the metal dopant with higher reactivity should be helpful.

In our case, to form a minimum (N=3) independent 4|4P-Se MTB loop, a critical density of niobium is needed, which is around 13 Nb dopants per nm$^2$ (cf. Eq. S1). In other words, it require ~5.5 Nb dopants to localize in 0.42 nm$^2$, which is the area of a minimum independent MTBs loop. By comparing the regions with different Nb concentrations, it is found that areal density of MTBs is proportional to the concentration of excess metal, similar to the Mo doping case[16]. Thus, we could engineer the areal density of MTBs by tuning the doping concentration, which provides an opportunity to form a dense and high-ordered MTBs networks also on WSe$_2$ where it has previously not been possible. Although we only tested Nb doping here, in principle other dopants like Mn, Fe, Co, Re, etc., may also work, as demonstrated previous in other 2D TMDC materials.[30, 33, 34]

Another interesting phenomenon in this study is that Nb dopants never appear in 4|4E-W MTB, which gives an opportunity to understand the formation mechanism of 4|4E-W MTB. We suggest that the transformation between 4|4E-W MTB and 4|4P-Se MTB are driven by the competition between strain and the bond between metal atoms. Generally, in the 4|4E-W MTB, Nb dopants will introduced a large stress due to the different lattice constants between WSe$_2$ and NbSe$_2$, which will push the neighboring metal atom out and form 4|4P-Se MTB to reduce strain, like the process in Fig. 4. On the other hand, unpaired electrons of W/Nb atoms will band with neighboring metal atoms in 4|4E-W MTB. Thus, the number of unpaired electrons will strongly influence the strength of band and the energy of system. For Nb and W in XSe$_2$ system (X is transition metal) the numbers of unpaired electrons are 1 and 2, respectively, which means that Nb dopants will lead a unstable 4|4E-W MTB.

## 4 Conclusion

To sum up, MBE grown Nb doped WSe$_2$ on HOPG has been studied at atomic scale via aberration corrected high resolution ADF-STEM. The high density MTB networks are successfully introduced by excess Nb doping. The 4|4P-Se MTBs present pseudo periodic features because of the random combination of W and Nb at the MTB, which provides an ideal specimen to study the confined 1D alloy systems in the future. We also report the first observation of a new type of MTB, which we denote as 4|4E-W. We propose a model to explain the formation and growth of MTB networks, which is supported by the first principles calculations. MTBs in TMDs materials could be introduced by excess metal doping due to the local metal rich condition in host crystals. Therefore, by choosing a suitable metal dopants and carefully controlling the growth parameters, MTBs can be engineered controllably even to those TMDs where it was deemed previously impossible, which can provide new opportunities for exploring novel material properties in 1D system.

## Methods

**Sample Preparation**. WSe$_2$ monolayers were grown via a multistep molecular-beam epitaxy (MBE) process as shown in Fig. 5. Step 1, WSe$_2$ monolayer were deposited on highly oriented pyrolytic graphite (HOPG) which was carried out in an Omicron UHV system with a background pressure <10$^{-10}$ Torr. The flux of W was generated from an e-beam evaporator and Se flux was provided from a conventional Knudsen cell, respectively. The flux ratio was kept Se/W ~20/1. The deposition temperature was 550 $^o$C with a growth rate of 0.25 MLsh$^{-1}$. Step 2, niobium flux from an e-beam heated source were introduced in to WSe$_2$ via a post-doping annealing process at 550 $^o$C for 24 min, during which the Se flux was not applied. Step 3, the doped sample was annealed at 550 $^o$C for 36 min to smoothen the sample during which the Nb and Se source were kept open that may lead to the growth of pure NbSe$_2$.

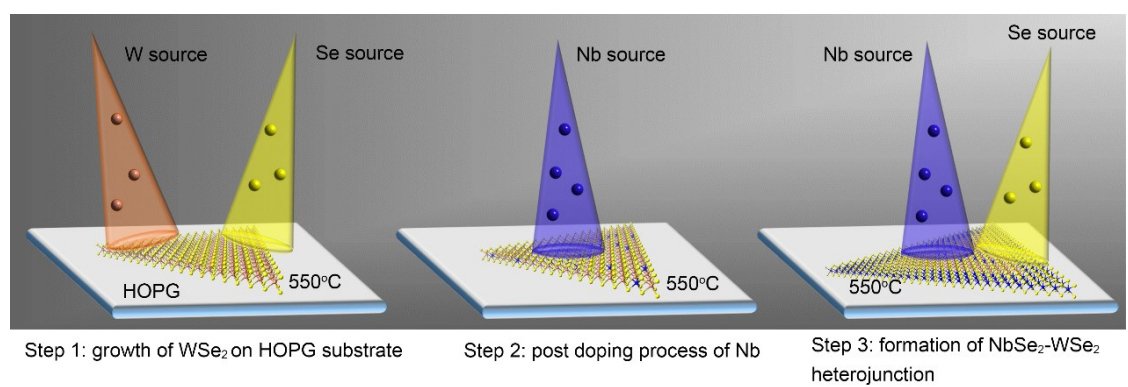

**Figure 5** MBE growth pathway of Nb doped WSe$_2$. Step 1, deposition of WSe$_2$ on HOPG. Step 2, introduce niobium sources doping niobium at 550°C. Step 3, introduce niobium sources and selenium sources to form NbSe$_2$ around the outmost Nb-WSe$_2$ and smoothen the surface.

**Characterization.** The MBE grown Nb-doping WSe$_2$ flake on HOPG was transferred onto a molybdenum TEM grid coated with lacey carbon film via a micromechanical exfoliation using paraffin wax as the coating and protection layer. To substantially decrease the contamination, the paraffin wax used was washed by acetone repeatedly. High-resolution ADF-STEM and selected area electron diffraction (SAED) were performed in a probe-corrected STEM (FEI Titan Chemi STEM) operated at 200 kV. The convergence angle was set at 21.4 mrad and the range of acceptance angle of ADF detector was 53-200 mrad, corresponding to strictly speaking, medium-angle annular dark-field STEM (MAADF-STEM)

**Image filter.** To enhance the contrast of 1D MTBs in relative low-mag ADF-STEM and hence map out their distribution in relative large scale, ADF-STEM images like Fig. 1(a) were treated via filtering out all the 1$^{st}$ and 2$^{nd}$ order pattern from their corresponding FFT spectrum. Followed by the inversed FFT, the resulting image was shown in Fig. 1(b).

**Calculation.** All calculations were carried out with the Vienna Ab Initio Simulation Package (VASP) [35, 36] using the Perdew-Burke-Ernzerhof (PBE) exchange-correlation functional [37]. The plane wave cutoff was set to 400 eV. The defects were modeled in a 10×10 supercell, with the Brillouin zone sampled using only the Γ point.

## Acknowledgements

The authors thank Zhoubin Yu for fruitful discussions, and Degong Ding, Haifeng Wang, and Daliang He for critical comments on editing the manuscript, and Dr. Fang Lin for kindly providing the codes for two-Gaussians filtering. B.W. and C. J. acknowledged financial support from the National Natural Science

Foundation of China under Grant Nos 51761165024, 51772265, and 61721005, the National Basic Research Program of China under Grant No. 2015CB921004, the Zhejiang Provincial Natural Science Foundation under Grant No. D19E020002, the 111 project (No. B16042) and Zhejiang University Education Foundation Global Partnership Fund. MHX acknowledges the financial support from a Collaborative Research Fund (C7036-17W) and a General Research Grant (No. 17327316) from the Research Grant Cuncil, Hong Kong special Administrative Region. We also acknowledge the support from the NSFC/RGC joint research grant (No. N_HKU732/17; 51761165024). HPK acknowledges financial support from the Academy of Finland through Project No. 311058, and CSC-IT Center for Science Ltd. for generous grants of computer time. The work on electron microscopy was carried out at the Center of Electron Microscopy of Zhejiang University.

# Niobium doping induced mirror twin boundaries in MBE grown WSe$_2$ monolayers

Bo Wang[1,2], Yipu Xia[3], Junqiu Zhang[3], Hannu-Pekka Komsa[4,*], Maohai Xie[3], Yong Peng[1], and Chuanhong Jin[2,1,5,*]

[1] *Key Laboratory for Magnetism and Magnetic Materials of Ministry of Education, School of Physical Science and Technology, Lanzhou University, Lanzhou 730000, China*
[2] *State Key Laboratory of Silicon Materials, School of Materials Science and Engineering, Zhejiang University, Hangzhou 310024, China*
[3] *Physics Department, The University of Hong Kong, Pokfulam Road, Hong Kong 999077, China*
[4] *Department of Applied Physics, Aalto University, 00076 Aalto, Finland*
[5] *Hunan Institute of Advanced Sensing and Information Technology, Xiangtan University, Xiangtan 411201, China*
*Email: Hannu-Pekka Komsa, hannu-pekka.komsa@aalto.fi; Chuanhong Jin, chhjin@zju.edu.cn

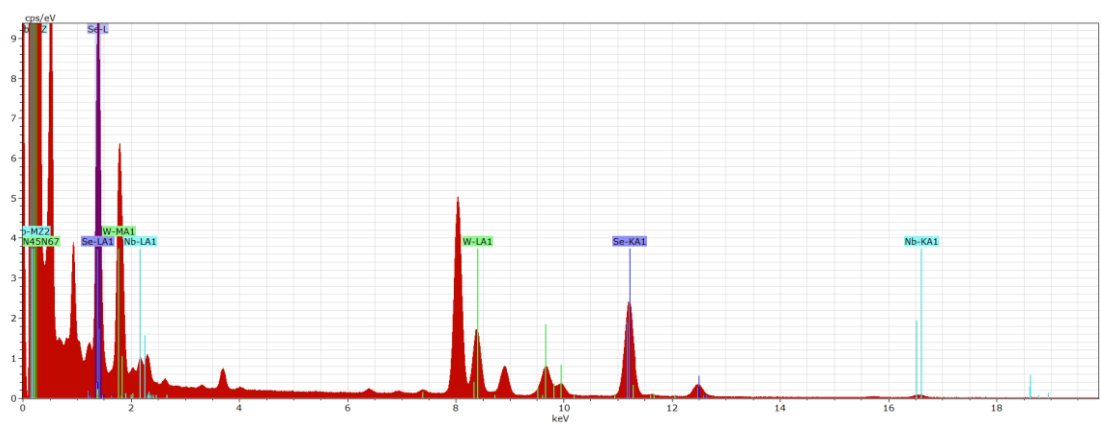

**Figure S1** Energy-dispersive X-ray spectrum (EDS) from Nb-doping WSe$_2$ sample. Characteristic signals for Nb, W, Se were clearly identified respectively.

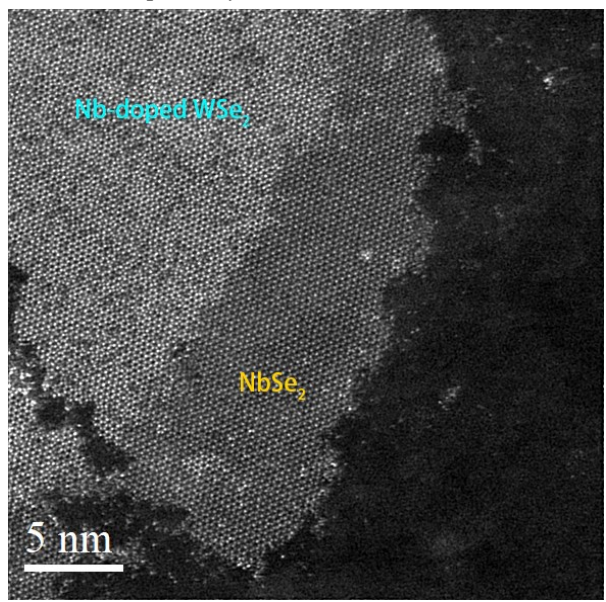

**Figure S2** ADF-STEM image of Nb-doped WSe$_2$ where MTB are not found in the NbSe$_2$ domain with slight W doping.

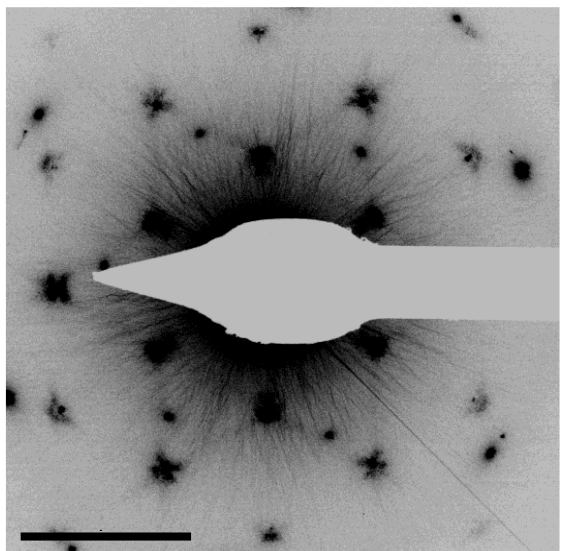

**Figure S3** SAED pattern of the Nb-doped WSe$_2$ domain shown in figure 1 (e). This pattern was non-linearly treated to improve visibility of David-star.

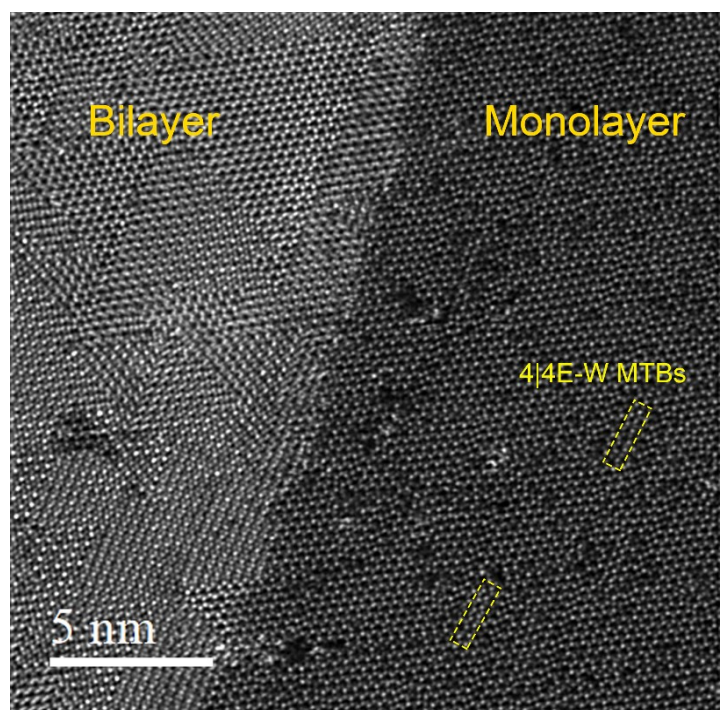

**Figure S4** ADF-STEM image of a Nb-doped WSe$_2$ region where 4|4E-W MTBs are indicated with dotted rectangle as shown in figure 3 (d).

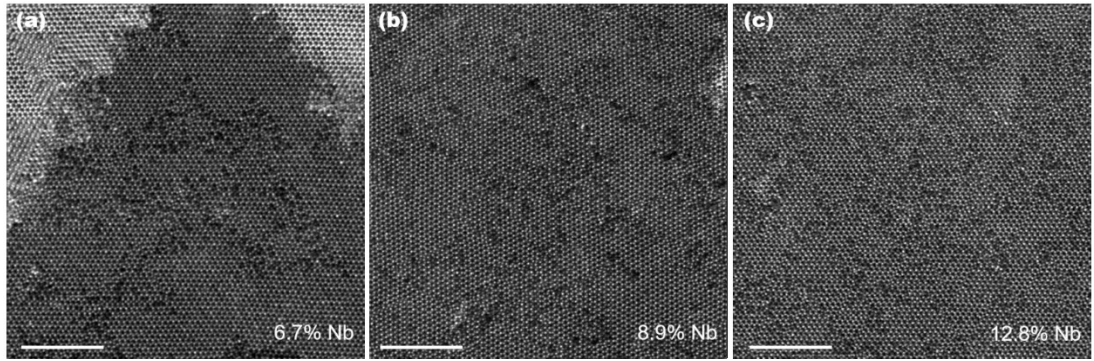

**Figure S5** (a-c) ADF-STEM image used for statistically analyzing the Nb concentration. Only monolayer regions were analysized for ensuring the statistical accuracy. Scale bar, 5nm.

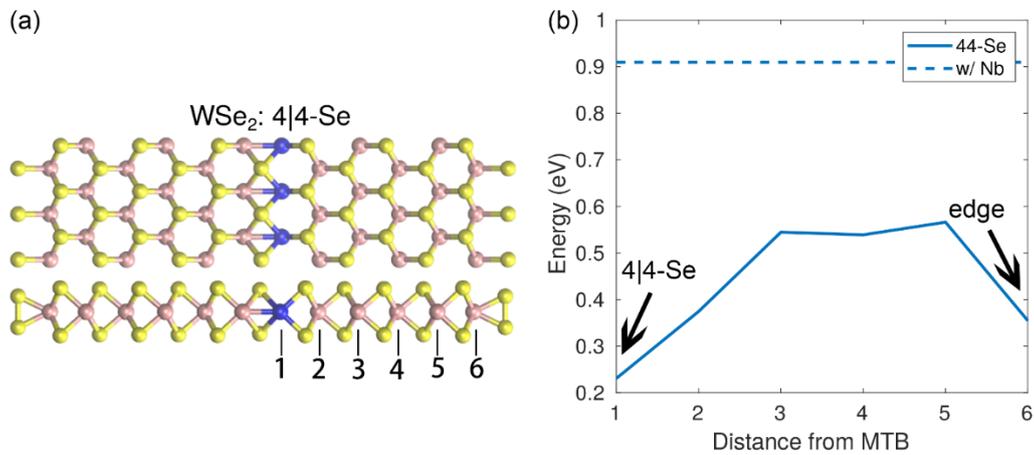

**Figure S6** Formation energy for Nb dopants at different sites in a WSe$_2$ ribbon with a 4|4-Se MTB located in the middle and Se2 passivated edges. (a) Illustration of the ribbon geometry used in the calculations, where the number represent the distance from MTB. (b) Formation energy of Nb dopants at different sites, it obviously shows Nb dopants prefer to stay at/around MTBs and edge of WSe$_2$ the lowest formation energy for site 1. The Brillouin zone was sampled using 12 k-points in the direction parallel to the MTB.

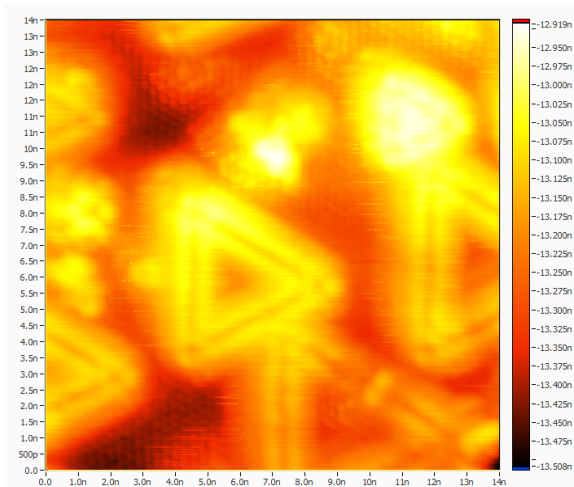

**Figure S7** STM image of Nb-doped WSe$_2$ without grow NbSe$_2$. Double-bright lines are the typical features of MTB.

The relationship between an independent $N_t$-size MTB loop and the local density of Nb it requires, could

be written as

$$D_{Nb} = \frac{N_t}{R} \frac{12\sqrt{3}}{N_t^2 a_{WSe_2}} \quad (S1)$$

where $D_{Nb}$ is the local density of Nb, $N_t$ is the number of tetragon in MTB, $R$ is the ratio of the number of tetragon to Nb, which is 1.63:1 as measured by experiment, $\frac{N_t}{R}$ is the number of Nb dopant needed to form a MTB who has $N_t$ tetragons, $\frac{12\sqrt{3}}{N_t^2 a_{WSe_2}}$ is the area enclosed by a MTB loop, and $a_{WSe_2}$ is the lattice constant of WSe$_2$. Thus, for $N_t$=9 (represent the minimum N=3 MTB loop), $D_{Nb}$ =13/nm², it gives a critical density of Nb to form an independent MTB seed in Nb-doped WSe$_2$. Furthermore, $D_{Nb}$ also represents the dispersive degree of Nb dopants. Eq.S1 shows that $D_{Nb} \propto \frac{1}{N_t}$, indicating that with the growth of MTB loop in size, the distribution of Nb would become dispersive, rather than clustering, which is also consistent with our experiment.